\documentclass[english]{IEEEtran}
\usepackage[T1]{fontenc}
\usepackage[utf8]{inputenc}
\usepackage{babel}
\usepackage{graphicx}
\usepackage[unicode=true]
 {hyperref}

\makeatletter

\usepackage{url}

\makeatother

\begin{document}

\title{Secure Information Sharing in an Industrial Internet of Things}

\author{Nils Ulltveit-Moe, Henrik Nergaard, László Erdödi, Terje Gjøsæter,
Erland Kolstad and Pål Berg\thanks{Nils Ulltveit-Moe, Henrik Nergaard, László Erdödi and Terje Gjøsæter
are with the Institute of Information and Communication Technology,
University of Agder, Jon Lilletuns Vei 9, 4878 Grimstad, Norway e-mail:
\protect\href{http://nils.ulltveit-moe@uia.no, henrik.nergaard@uia.no, laszlo.erdodi@uia.no, terje.gjosater@uia.no}{nils.ulltveit-moe@uia.no, henrik.nergaard@uia.no, laszlo.erdodi@uia.no, terje.gjosater@uia.no}.}\thanks{Pål Berg is with Applica Consulting, Postboks 113, Rådhusveien, 4524
Lindesnes, Norway, e-mail: \protect\href{http://pal.berg@applica.no}{pal.berg@applica.no}.}\thanks{Erland Kolstad is Advisor at Devoteam AS, Jon Lilletuns vei 1, 4879
Grimstad, Norway e-mail: \protect\href{http://erland.kolstad@devoteam.no}{erland.kolstad@devoteam.no}.}}
\maketitle
\begin{abstract}
This paper investigates how secure information sharing with external
vendors can be achieved in an Industrial Internet of Things (IIoT).
It also identifies necessary security requirements for secure information
sharing based on identified security challenges stated by the industry.
The paper then proposes a roadmap for improving security in IIoT which
investigates both short-term and long-term solutions for protecting
IIoT devices. The short-term solution is mainly based on integrating
existing good practices. The paper also outlines a long term solution
for protecting IIoT devices with fine-grained access control for sharing
data between external entities that would support cloud-based data
storage.\end{abstract}


\IEEEpeerreviewmaketitle{}

\section{Introduction}

\IEEEPARstart{T}{he} Industrial Internet describes industrial processes
controlled by SCADA systems and similar that are being networked and
interconnected across the value chain to create smart integrated production
systems. The Industrial Internet phenomenon embraces the Internet
of Things (IoT) domain, where smart production based on RFID tagged
products and sensor networks are being integrated into an Industrial
Internet of Things (IIoT). This allows for improving the quality,
traceability and integrity of industrial processes by allowing better
modeling of the Cyber-Physical Systems (CPS) and processes using techniques
such as data mining, big data analysis, learning systems and knowledge-based
systems using semantic modeling and ontologies.

It also improves maintainability, reliability and availability of
the controlled industrial processes by using sensor networks for Condition-Based
Maintenance (CBM) in order to monitor wear and pre-failure of technical
components. This reduces the risk of system breakdowns during production,
and allows for planned exchange or upgrade of production equipment
with lower risk of excessive downtime during repairs.

The most important actors considered in this article are:
\begin{itemize}
\item Industrial organisations that want to introduce Internet of Things
(new markets).
\item Industrial organisations that require secure exchange of information
related to IoT or need information related to their supplied equipment.
\item Third party organisations that require limited access to some of the
sensor data, for example vendors providing sensor or equipment maintenance
or managed security service operator equipment. 
\end{itemize}
An important objective for manufacturers is to improve the production
efficiency whilst reducing planned and unexpected downtime. IoT may
help to achieve this goal, however in order to do this, any efficiency
improvements must be measurable. The Overall Equipment Effectiveness
(OEE) is a well-known metric of manufacturing efficiency that can
be used for this. It is defined in terms of the availability $A$,
performance efficiency $P$ and quality rate $Q$ as $OEE=A\cdot P\cdot Q$. 

A problem with OEE, is that it in itself is not sufficient, since
it only provides the status of production efficiency, and blurs the
relationship between performance and cost involved in sustaining a
given OEE level~\cite{lee_recent_2013}. It does for example not
show the relationship between invisible pre-failure and wear conditions
and the production performance. Furthermore, when a device eventually
has failed, possibly interrupting production, then this will already
have caused a loss in production efficiency. 

Data mining from condition-based maintenance monitoring sensors is
therefore one area where production companies can improve productivity
beyond what OEE easily can measure. This allows for performing predictive
fault analysis and control functions in order to provide more resilient
effectiveness~\cite{lee_recent_2013}. 

\begin{figure}
\includegraphics[scale=0.42]{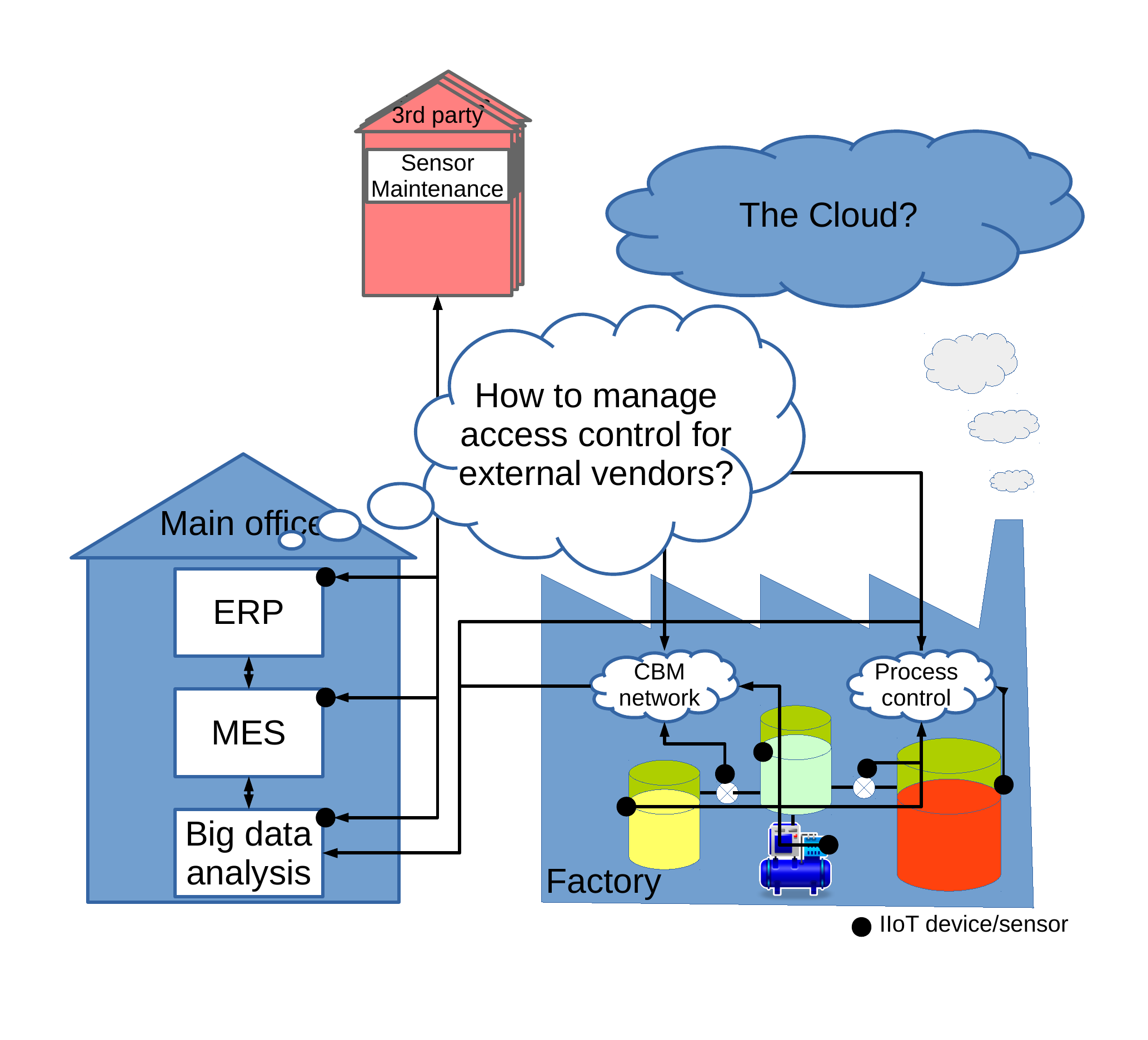}

\protect\caption{\label{fig:High-level-figure-illustrating}High-level figure illustrating
the problem with data sharing of sensor data for industrial processes.}
\end{figure}

IoT systems helps in laying the foundations for such predictive manufacturing
by providing the essential structure of smart sensor networks and
smart machines~\cite{lee_recent_2013}. Communication protocols,
such as the Object Linked Embedded for Process Control, Uniformed
Architecture (OPC-UA) facilitate platform independent data acquisition
from these sensors using a Service Oriented Architecture (SOA) based
on web services~\cite{hannelius_roadmap_2008}. OPC-UA also supports
vertical integration between different layers of factory automation,
such as Enterprise Resource and Planning (ERP) systems for factories,
Manufacturing Execution Systems (MES) and automation systems~\cite{hannelius_roadmap_2008}.
It even supports integrating data with systems in partner companies,
as illustrated in Figure \ref{fig:High-level-figure-illustrating}.
Security, reliability and AAA (Authentication, Authorisation and Accounting)
are also integrated into the OPC-UA standard, which supports an API
mapping to XML web services focusing on interoperability as well as
an UA native mapping focusing on efficient low-bandwidth data transfers~\cite{hannelius_roadmap_2008}. 

The core characteristics that typically identifies IoT devices, such
as smart sensors are: 
\begin{itemize}
\item Interaction with the physical world 
\item Have communication capabilities (device to person, devices to device,
and device to multiple devices) 
\item Have some processing capabilities (e.g. support decision making)
\end{itemize}
The market for IoTs is believed to be rapidly increasing\footnote{In 2020, 25 billion connected things in use: http://www.gartner.com/newsroom/id/2905717},
as services utilising widely deployed sensor networks become generally
available, such as smart home equipment, smart cars etc. at the same
time as any device now typically will have the capability of being
networked. This means that there will be a large amount of mass-produced
and affordable sensor technologies that can be deployed everywhere,
including in industries.

This means that the production equipment in manufacturing is becoming
more and more advanced and smarter by supporting inherent abilities
for decision making. However, both operation and maintenance require
expert knowledge, and often it will be external parties that have
this knowledge. There will therefore be a need for \emph{secure information
sharing and collaboration among stakeholders}, as illustrated in Figure
\ref{fig:High-level-figure-illustrating}. 

The figure shows that the company often will collaborate with several
third parties, for example external vendors or entities who will get
access to some data from industrial IoT devices inside the company.
Examples of such devices are: \emph{intrusion detection appliances}
as part of a managed security service; \emph{vendors}, \emph{suppliers}
or \emph{trusted third parties} managing custom sensors for monitoring
vibration, wear or temperature for Condition-Based Maintenance (CBM)
as well as \emph{flow}, \emph{temperature}, \emph{pressure} or \emph{position}
in IIoT devices controlling industrial process etc. 

The main problem is that all these actors need access to \emph{their}
devices, but should only have access to these according to a strict
definition of need, ensuring minimal spillover of other company sensitive
information about production processes etc. as possible. Furthermore,
not only the device, but also the data and information generated by
the device will need to have constraints in the form of detailed access
control, especially in multi-sensor devices.

The scenario furthermore shows that the company is performing data
analysis, potentially using big data from several manufacturing plants,
in order to extract necessary indicators on production quality while
also analysing pre-failure and wear based on the CBM sensors. Another
example is analysing for signs of cyber-attacks on either the corporate
or process network. External vendors may then be notified if anomalous
data are detected, so that these then can do further troubleshooting
via an interface towards the sensors they are authorised to manage.

There is also a security and safety risk with such external parties,
since it increases the amount of people who will have potentially
deep access into industrial control networks. This increases the risk
of someone disrupting industrial processes if they have malicious
intent. Not the least will there be a push towards outsourcing internal
services, for example ERP systems, to cloud service providers, which
adds additional challenges when it comes to managing the services
securely, as well as reducing the risk for leakage of corporate private
material. 

Protecting data transmissions in a secure manner is technically relatively
easy to achieve using existing and standardised cryptographic methods
such as Transport Layer Security/Secure Sockets Layer (TLS/SSL) or
Datagram Transport Layer Security (DTLS). However a challenge is
managing and provisioning keys and digital certificates as well as
handling user and service authorisation in a scalable and manageable
way.

The main problem is that current solutions for managing access in
general are too coarse-grained. Firewall rules will for example give
access to the entire CBM or process control network without limiting
access to the sensors and the data that external parties are allowed
to access, especially when the data is mixed.

This paper discusses how security can be improved in industries wanting
to utilize the power of IoT, especially focusing on how different
stakeholders such as customers, subcontractors and equipment suppliers
can be granted access to sensor data and other data from the manufacturing
process in a controlled and secure manner, without compromising sensitive
data that are not shared.

The rest of this paper is organised in the following manner: Section
2 presents the background of industrial IoT, while their requirements
are further explained in section 3. The current challenges is explained
in section 4. Section 5 presents relevant cases from the industry
and section 6 gives an insight into relevant standards. We present
a short term solution that can improve the current security in section
7, while section 8 elaborates on the possible long term solutions
which could be applied for a more secure and trustworthy IoT.  Current
existing research is presented in section 10, Related Work. Section
11 discusses and summaries the paper, while the last section; section
12 gives an overview towards possible future work.

\begin{figure}
\includegraphics[scale=0.38]{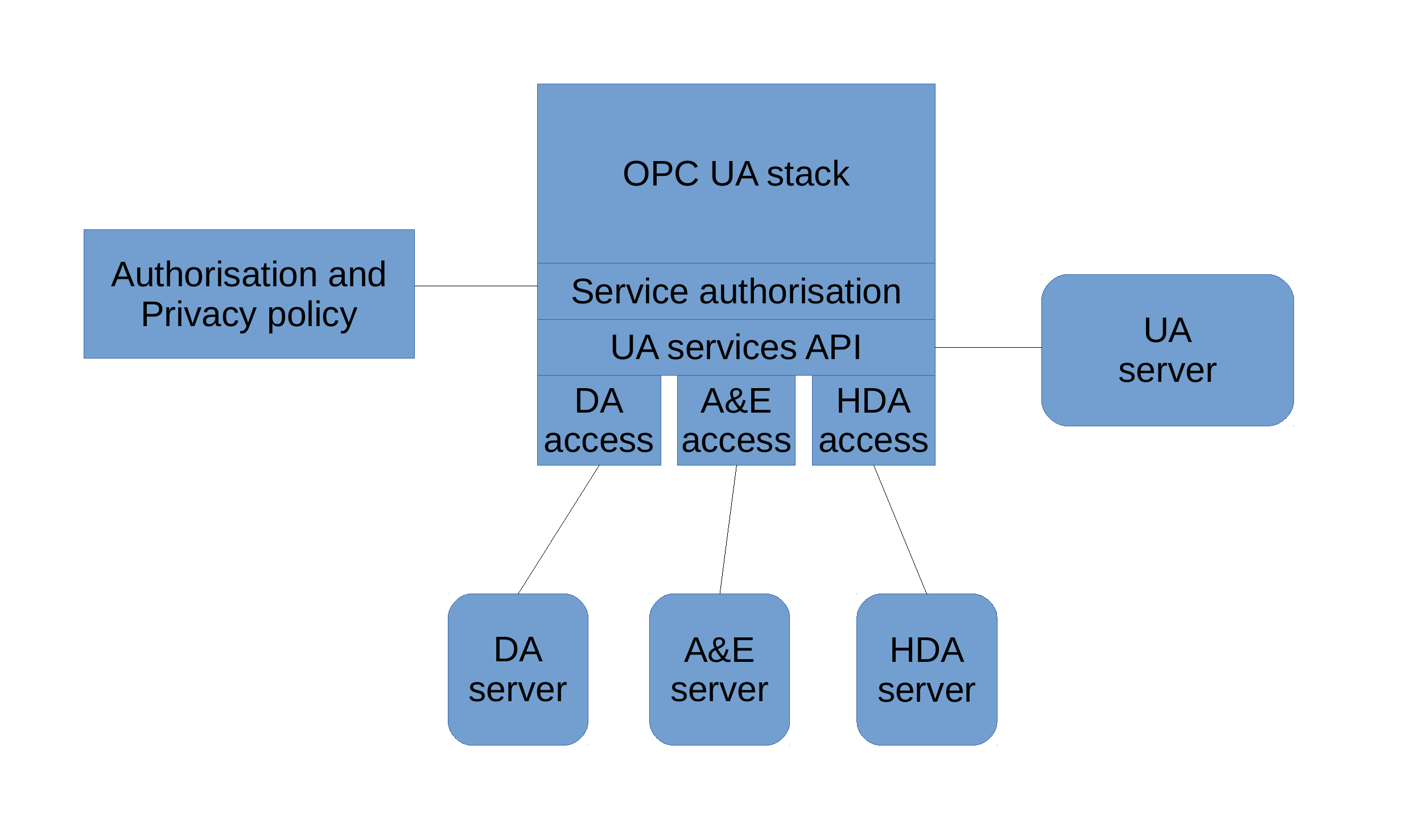}

\protect\caption{\label{fig:OPC-UA-gateway}OPC UA gateway supporting fine-grained
access control to OPC UA services.}
\end{figure}

\section{Background}

New technology and availability of affordable mass-produced sensors
and devices enable new possibilities to meet requirements for continuous
improvements of productivity and efficiency.

The industry looks at IoT and digitalisation of sensors as part of
Industry 4.0 and sees them as tools to improve productivity and reduce
costs, e.g. through CBM. Low-cost sensors with good enough precision
and lifetime enable more detailed process insights, and better process
optimization.

With the ever increasing capabilities of computing hardware, new technology
will cost less and become more readily available. Small sensors with
the ability to be interconnected into the Internet, enables a plurality
of new possibilities. In an industry setting, this could provide a
reasonable and efficient way to gather more information of the production
process giving new opportunities for optimisation. Easy sensor deployment
increases the potential for cloud based data mining and analytics
using big data from semantic sensor networks, virtual sensors and
complex event processing~\cite{vermesan_research_2014}.

\subsection{The Industry 4.0 Challenge}

The term industry 4.0 (Industrie 4.0) is a German strategic initiative
for strengthening the competitiveness of the German manufacturing
industry based on an association of representatives from business,
academia and politics \cite{kagermann_industrie_2011}. Similar ideas
have approached also outside the German area, such as Industrial Internet,
Advanced Manufacturing, Integrated Industry and Smart Industry \cite{hermann_design_2015}.

There has so far not been any clear definition of what Industry 4.0
is, however a meta-study of 200 publications describing the concept
was done by Hermann et al., who came up with the following definition
of Industry 4.0 \cite{hermann_design_2015}: \emph{Industrie 4.0 is
a collective term for technologies and concepts of value chain organization.
Within the modular structured Smart Factories of Industrie 4.0, CPS
monitor physical processes, create a virtual copy of the physical
world and make decentralized decisions. Over the IoT, CPS communicate
and cooperate with each other and humans in real time. Via the Internet
of Services (IoS), both internal and cross-organizational services
are offered and utilized by participants of the value chain.}

Security is a challenge with IIoT and Industry 4.0, where heavily
interconnected production systems exchange information and data, not
only within the manufacturing facility, but also across the value
chain to corporate Enterprise Resource and Planning (ERP) systems,
customers, subcontractors and equipment suppliers. One large problem
with IIoT, is that it is integrated with the control systems of existing
production facilities which may have a lifetime of decades and was
originally built without security or Internet connectivity in mind. 

The security of control system protocols have also lagged behind the
security of information technology (IT) systems, and is only now starting
to get more widespread use. This means that many devices, which were
never intended to be networked, may be interconnected in an IIoT setting.
This creates a huge attack surface towards devices that may not be
able to protect neither the data integrity nor data confidentiality
as well as frequently having weak access control mechanisms, like
requiring default user names or passwords~\cite{crysys_lab_duqu:_2011,lab_duqu:_2011}.
Cars or airplane systems are examples of such real-time systems where
critical systems often share the same information bus, which makes
it highly dangerous if one device has malicious behavior.

\section{Industrial Internet of Things Requirements}

There are several important requirements that an infrastructure handling
industrial IoT devices must fulfill.

\subsection{Real-time data transfer}

Control systems typically require timely delivery of information.
What is consider real-time depends on the process being controlled.
The inner process control loop may need to control processes down
to millisecond precision without any loss of control signals. This
means that the process control network will have very limited tolerance
for variations in latency which causes problems when using traditional
Internet security protocols such as TLS/SSL. SSL key renegotiation
would for example cause large problems for such a process, and even
running such a process over TCP/IP might not be feasible. Other processes
have less strict real time requirements, and will be able to run over
traditional Internet links without problems.

\subsection{Availability}

It is typically a basic requirement that information is always available
and accessible to authorised users and services. This is also emphasised
in the OEE metric, where availability is one of the foundational metrics
of service quality. Availability also implies data persistence, i.e.
ensuring that data does not suddenly disappear due to failure (disks
wearing out, lightening strikes etc). Another concern may be legal
issues causing obstacles for data availability between countries,
as well as a concern that foreign authorities may unrightfully gain
access to company sensitive information.

As industry moves towards an IoT scenario, then there will be a requirement
that these data are available from everywhere and to everywhere. Data
must be available both between production facilities, device suppliers
as well as subcontractors and the users themselves, who will expect
to be able to purchase tailor-made industrial products. The car industry
is already at the front of such production by providing tailor made
products according to the customers wishes. This again means that
industrial data needs to be made available also via cloud-based services
in order to provide the necessary scalability to handle a large customer
base or for reducing the operational costs of managing IT equipment.

\subsection{Secure information sharing }

Secure information sharing implies that there exists some data that
can be shared with partner organisations, or between daughter companies,
while preserving data confidentiality and integrity. Existing cryptographic
building blocks, such as public key encryption, symmetric encryption,
message authentication codes etc. can be used for enforcing this.
One of the main challenges in secure information sharing is scalable
solutions for handling identities and authorizations, including protocols
for managing keys, encryption protocol upgrades and digital certificates
 This is also where many IoT protocols (e.g. Zigbee, ZWave etc.)
have been shown to be flawed~\cite{dini_considerations_2010,fouladi_honey_2013}.

Secure information sharing includes non repudiation, so that partner
organisations cannot deny having done certain operations. The latter
can for example be implemented using secure logging schemes~\cite{bezzi_secure_2010,ulltveit-moe_novel_2014}.

\subsection{Information Leakage Detection and IPR-handling}

Preventing information leakage includes data leakage detection and
IPR handling, for example detecting whether process sensitive information
is leaked from the owning industry, and found stored in inappropriate
places. A possible solution can be using Digital Rights Management
(DRM) type of technologies to limit the possibilities of data leakage
by strong cryptographic access control methods to the information,
as well disallowing copy/paste of this information between a trusted
and untrusted application.  DRM can be tied to hardware, like the
Trusted Platform Module (TPM). There are already scalable solutions
for decrypting quite high bandwidths today, e.g. satellite HD video
etc. Limiting data access can also be done using more traditional
techniques, such as limiting data access using dumb terminal servers
(e.g. Citrix servers) allowing only limited access to sensitive data
inside the production plant. A challenge in both of these cases, is
that some data leakage still may occur, for example by taking screenshots
of the terminal window or software with DRM protection, or even taking
digital photographs of the screen used to present these data using
external devices (cameras, mobile phones etc.). The information owner
will therefore need to trust the external parties to some extent,
however it is possible to limit the possibilities for other types
of data analysis and data correlation than the data owner desires
using such measures.

Challenges with real-time data due to network latency may be a problem
in some use scenarios, however other use cases are less time critical
given the latencies of encrypted traffic on the Internet.

Techniques such as digital watermarking or tagging of information
can be used to enforce nonrepudiation for such data leakages~\cite{papadimitriou_data_2011}.
It is however questionable how useful digital watermarking of sensor
signals will be for sensor data, since this adds noise which may interfere
with the signal quality. Another technique that has been proposed
is entropy-based metrics for detecting information leakages and verifying
security policies, in order to detect accidental information leakages
due to faulty security or privacy policies~\cite{ulltveit-moe_measuring_2013}.

\subsection{Flexible production}

Flexible production is at the core of the Industry 4.0 vision and
implies a requirement for reconfigurability of production cells within
the industry, so that these easily can be repurposed and assigned
to other product lines on demand according to purchase orders. This
implies that there must be tight integration between the ERP, MES
and factory automation system, so that production cells can be reprogrammed,
moved and assigned to the product lines where they are most urgently
needed, without compromising the logistics of raw materials, dependent
products and finished products.

\subsection{Decision support}

Decision support systems can be used both for planning the production,
for condition-based maintenance as well as for handling logistics.
Information can be mined using different data mining techniques such
as data warehousing or big data analysis in combination with artificial
intelligence or learning systems. Another component that frequently
is used with decision support systems is ontology based reasoners
that are able to infer new knowledge based on information stored in
the ontology~\cite{thomalla_ontologie-basierte_2014}. Decision support
is traditionally done in-house, but can also be done distributed based
on data in the cloud, for example to measure customers opinions towards
the company's products.

 CBM is a typical example where third parties can have access to
analysing and monitoring facilities, and can have means for requesting
shutdown of equipment based on condition data.

\subsection{Fine-grained Access Control to Data}

Sophisticated access control mechanisms is in particular relevant
when data is shared among multiple parties and come from a variety
of sources, which is the case for typical IIoT scenarios. It has been
suggested that a transition from a traditional Role-Based Access Control
(RBAC) infrastructure to a more fine-grained Attribute-Based Access
Control would be required in order to manage access to an IoT based
infrastructure~\cite{vermesan_research_2014}. Attribute-based access
control mechanisms, such as the eXtensible Access Control Markup Language
(XACML)~\cite{moses_oasis_2005}, has for example been proposed used
in IoT gateways by the EU FP6 integrated project SANY (Sensors Anywhere\footnote{SANY project: http://www.opengeospatial.org/ogc/regions/SANY}).
This project implemented an open web service based architecture for
sensor networks~\cite{uslander_designing_2010}. SANY was based on
the outcome from the EU FP6 project ORCHESTRA, which provided a specification
framework for the design of geospatial service-oriented architectures
and service networks~\cite{open_geospatial_consortium_rm_oa_2007},
as well as a test bed for implementing aspects of the Geographical
DRM reference model (GeoDRM)~\cite{open_geospatial_consortium_geodrm_2006}.
SANY is implemented as a network proxy providing a quick and cost-efficient
way to reuse data and services from currently incompatible sensors
and data sources aimed at environmental monitoring. The SANY project
also did the initial prototype of the GeoXACML OGC standard~\cite{ed_ogc_2007,matheus_geospatial_2008},
which provides geographical access control to sensor devices, based
on geographic locations defined by the Geography Markup Language version
2 (GML2)~\cite{ed_ogc_2002}. 

An interesting feature with this architecture, is that it acknowledges
that an IoT architecture will be based on geographically dispersed
sensors, which also means that geographic access control based on
advanced geographic information system (GIS) primitives will be needed
in order to manage access to these sensors.

RBAC and ABAC are standard technologies, however they cannot be yet
applied in a straightforward way to IIoT scenarios. The SANY web service
architecture does for example not support OPC-UA. However there are
research efforts that in the long run may mitigate this. The PRECYSE
project developed and used the Reversible anonymiser~\cite{ulltveit-moe_novel_2014},
which enables anonymising messages, using a policy based approach
for specifying which parts that should be anonymised. It furthermore
has the ability to de-anonymise these parts again for authenticated
and authorised users. This solution is under further development in
the currently running SEMIAH project, together with a graphical block
based tool to construct and mange the polices used by the Reversible
Anonymiser~\cite{nergaard_scratch-based_2015,gjosaeter_security_2014}.
\emph{}

\emph{}

\section{Challenges}

This section describes the main challenges that may occur when attempting
to share data in an industrial internet of things scenario. The obvious
challenge in an industrial setting, focusing on utilising IIoT for
automation, and analytical gains is how to enssure that the introduced
things are secure and tamper proof. This includes discussing the how
IIoT devices impacts the attack surface. Subsequent sections describe
how to best protect and prevent these attacks on an industrial internet
of things, among others by applying a defense in depth strategy.

The main challenge is how to balance the need for security on one
side with the ability to share and utilize the possibilities in IIoT
on the other side. What is sufficient and secure enough?  

Security is not achieved by only implementing secure devices or encryption
of information. In any system, including IIoT there are some key components
that are essential to achieve a secure solution;
\begin{itemize}
\item Secure device
\item Access control, including identity management, authentication and
authorization
\item Secure communication
\item Management, and
\item Trust
\end{itemize}
Access control can be enforced at several layers, should it be on
the network, device or data layer? Access control on the data layer
will give the most flexible solution, but also represent the most
complex authorization scheme.

Many of the industrial sensors are furthermore resource constrained
devices running real-time processes with limited processing capability,
which means that traditional software security mechanisms, such as
using a public key infrastructure with standard encryption mechanisms
may not work due to unacceptable latencies or lack of processing capacity
for example during key renegotiation. Devices tend to be made to run
on exactly the minimal required hardware specification so there is
little to no leverage to add security components.

Another security challenge is that device manufacturers have a bad
track record when it comes to adding backdoor capabilities to their
devices in order to manage or update these devices (for example \cite{goodin_malicious_2015}).
The original intent for installing such backdoors may be valid enough,
however the problem is that these typically use a very simple security
solution - often only using a standard username/password, which obviously
is not secure in today's Internet. It is therefore important that
the devices themselves also can be integrated into the organisations'
AAA infrastructure.

\subsection{Legacy-systems}

One problem in industrial systems occurs when upgrading to a modern
IIoT from an older system which did not account for global connectivity.
This can lead to security issues, especially when some of the devices
have not considered the possibility of appearing in a non-restricted
network, when part of the control network is being bridged to other
networks, which in turn may be exposed to the Internet. Another problem
is that legacy systems tend to have non-differentiated networks and
coarse-grained access control if any.

\subsection{Threats}

The threat landscape is quite different across different industries
and sectors. In some industries the secrecy of the processing methods
are considered essential for the company existence, while other industries
have completely different issues. Industries with high value IPR's
are also exposed to more advanced threat actors than industries with
less IPR\footnote{FireEye Annual Threat Report: https://www.fireeye.com/current-threats/annual-threat-report.html}.

The insider threat is perhaps one of the largest type of threats in
an IIoT scenario, since own employees as well as vendors and others
may be authorised to access and/or control sensors in the network.
New functionality may be added to the control network, such as remote
support or upgrading from employees home or from partners, desire
to get data/statistics from production. Need a risk assessment when
implementing such solutions. Another challenge is that suppliers may
have their own links towards their systems. If these stop, then this
may even stop the factory. 

It is important that fine-grained access to the entire sensor network
can be managed centrally by the network owner, so that access can
be granted or denied quickly and precisely to specific devices. Certain
operations, such as configuration deployment, should also support
multi-party authorisation policies for example based on key shares~\cite{shamir_how_1979,ulltveit-moe_novel_2014}.
This reduces the risk that corrupt or radicalised insiders are able
to destabilise the factory infrastructure by deploying malicious or
faulty system configurations. There will for example be a significant
insider threat if laid off employees, or external parties with terminated
contract still have access to the system. Another example is that
the owner may get sensors installed on the factory premises which
communicate with external parties using mobile communications, for
example GPRS, where the network owner is not aware of what information
the external party is able to extract using these managed sensors.
This is a significant concern, since such sensors often are based
on generic multi-sensor gateway platforms running traditional operating
systems (often Linux) which may be able to communicate using many
wireless different protocols in addition to the use the sensor is
intended for. This means that a malicious or hacked third party device
potentially would be able to compromise internal wireless sensor networks
on the production facilities.

\subsection{Security Attacks on an IIoT}

According to the Jupiter research 38.5 billion IoT devices will be
on the planet by the 2020 \cite{Jupiterresearch}. These devices will
mainly be smart phones, smart house devices, e-health devices and
cars, but there will also several unique devices for specific proposes
(e.g. watches, glasses, body analyzers, etc). As the number of the
IoT devices proliferate, the challenges for security professionals
in the form of attack surface and attack types will increase, perhaps
to the level where these problems become unmanageable. With this tendency,
the protection of these devices will be an extremely important and
difficult task.

The number of cyber-attacks shows a concerning tendency. The number
of cyber-attacks is expected to be doubled between 2011 and 2017 \cite{Cybercrimestatistics}
\cite{McAffeeThreatReport}. This tendency predicts an even higher
growth for IoT devices in the future, because of the huge spread of
such devices \cite{AnatomyofAttack}. Attacks can aim at stealing
personal information, gaining money, etc. The attacks are also able
to intervene with the normal operation and cause unavailability, annoyance
or damage, or they can be used for preparing further synchronized
attacks.

The most dangerous attacks are based on zero day vulnerabilities,
which are formerly unknown attacks that typically will go undetected
by anti-malware software. In this case the window of exposure and
overall impact can be extremely high. If an unknown software error
appears, then millions of IoT devices can become vulnerable instantly.
Several cases has been detected when critical errors were found in
crucial software components (web server application, encryption weaknesses,
compression software tools etc.) \cite{HeartBleed} \cite{LZOVulnerability}.
The problem escalates if the new vulnerability is not patched immediately,
so that common exploits appear to take advantage of the vulnerability.

The number of zero day vulnerabilities is expected to be around 700
by the year of 2015 but it also shows increasing tendency \cite{CVEDetails}.
Apart from zero day errors there are other serious threats which are
related to configuration errors, improper usage of tools and also
to the human factor. The following figure shows the main attack surfaces
of an average IoT device. 

\begin{figure}
\includegraphics[scale=0.38]{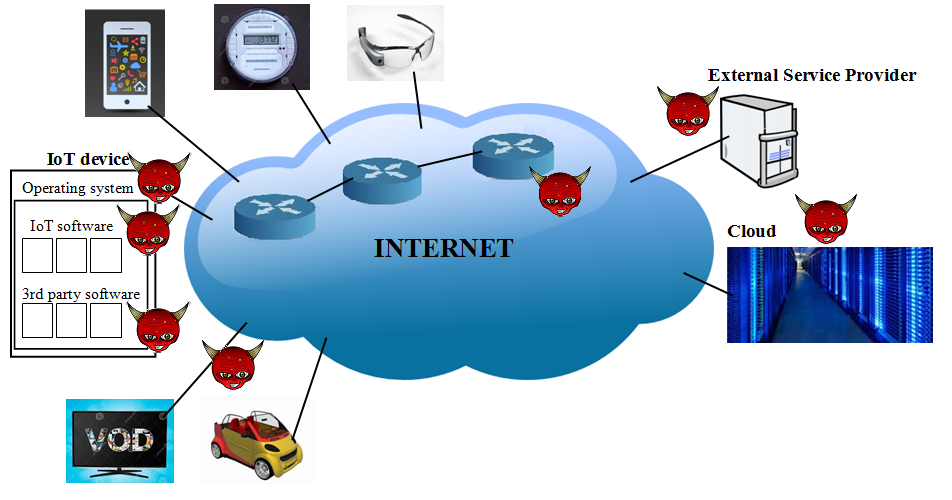}

\protect\caption{\label{fig:OPC-UA-gateway-1}General overview of the attack surface
of the Internet of Things.}
\end{figure}

For any IoT device the following relevant attack surfaces can be mentioned:

\begin{itemize}
\item IoT device operating system:

\begin{itemize}
\item configuration error (e.g. unnecessary services, factory default passwords)
\item software error (e.g. lack of input data validation, memory corruption)
\end{itemize}
\item IoT device own software

\begin{itemize}
\item configuration error (e.g. weak authentication method, lack of protection
against denial of service)
\item software error (e.g. arbitrary code execution though API)
\end{itemize}
\item IoT device 3rd party software

\begin{itemize}
\item configuration error (e.g. lack of encryption, parameter tampering)
\item software error (e.g. file inclusion vulnerability)
\end{itemize}
\item error in the communication channel (lack of encryption, man in the
middle, cryptographic weaknesses)
\item vulnerability in the internal network devices (e.g. information disclosure,
traffic poisoning)
\item vulnerability in the external individual service provider (all kind
of web service, database service vulnerabilities) 
\item vulnerability in the cloud service providers 
\end{itemize}
As the number of vulnerabilities increases, the purpose of an attack
has been extended during the years (stealing personal information,
causing annoyance and anger, causing damage, sophisticated spying,
cyber terrorism and even cyber war).

The following types of adversary objectives are the most relevant
\cite{KasperskySecBullet2013}:
\begin{itemize}
\item information leakage (stealing information e.g. health data, habits)
\item stealing money (attacking the bank transfer service of an IoT)
\item integrity changing (modifying data for the attacker's benefit or causing
annoyance)
\item damaging reputation (attacking successful companies or high traffic
service providers)
\item availability related: wiping data or blocking operation (causing denial
of service can be critical e.g. in the energy sector)
\item sophisticated attacks (malware, attacks through command and control
servers, etc.) 
\item cyber terrorism (IoT devices can be connected to critical infrastructures) 
\end{itemize}
A specific IoT device will typically have a specific attack surface.
Several IoT related vulnerability was detected and analyzed during
the last years. An internet connected gun is analyzed and unauthenticated
API and short guessable PIN is detected in 2015 \cite{RifleHack}.
A vulnerability in the firmware of a network device was revealed which
would expose millions of IoT devices to an attack~\cite{VectraThreat}.
A baby monitoring device vulnerability can cause lot of annoyance
to its users \cite{BabyMonitor}, etc.

The attack on the Internet of Things can be more dangerous and can
have more critical effects if the targeted computer is an industrial
machine. Several virus attack was detected against SCADA systems during
the last decade. The Slammer virus \cite{Slammer} targeted Nuclear
Power Stations in the USA in 2003, Conficker \cite{Confiker} has
several targets including navy systems as well. The first very sophisticated
malware that was detected for such purposes was Stuxnet \cite{Stuxnet}.
Stuxnet specifically targeted Programmable Logic Controllers (PLC)
of centrifuges for separating nuclear material. Stuxnet was designed
to infect modern SCADA systems as well as PLCs. A very similar malware
named Duqu \cite{crysys_lab_duqu:_2011} was discovered later which
aimed to collect information for further Stuxnet like attacks. Stuxnet
has several variants, and probably belongs to the same root such as
the Secret Twin of Stuxnet \cite{StuxnetTwin} or the Flame \cite{Flame}.
There are several cases when a malware is customized to specifically
target industrial IoT. A variant of the Havex malware targeted industrial
control system and SCADA users in the middle of 2014~\cite{Havex}.
Because malware variants appear very rapidly and can be customized
for specific architectures and tasks, it is clear that Industrial
IoT hardly differ as a target, but the societal effect of a successful
attack can be much higher than attacks on traditional consumer-oriented
IoT devices.

\section{Relevant Industry Cases}

Our main focus has been the process industry and technologies around
integrated operations in oil and gas. As mentioned initially in this
paper, the process industry is one of the industries for which the
concepts of Industry 4.0 will be very relevant. In our survey where
we interviewed managers and persons in charge of cyber security, IoTs
were one of their main concerns. Manufacturers of equipment and third
party service providers wanted to have access to their equipment and
sensors or IoT devices, which were internal to the plant's network.
They see IoT as beneficial to both cost and quality, but struggles
to have a security strategy which incorporates this new paradigm.

Information sharing was not the biggest concern, as data from IoT
devices often were specific for the equipment and revealed little
secret information about the manufacturing process. However, if increasingly
more devices are installed, then external parties will get a better
understanding of the industrial processes which is not acceptable.
Data sharing was performed by using traditional methods like VPN with
username and password as credentials, firewall routing and role based
access control. It could be initiated by the external party after
the initial registration and configuration processes were finished.
Data quality is a concern, but this will be discussed in our use case
for the oil and gas industry. Lastly, security of the IoT device itself
with resistance to attacks and hostility was a challenge as hacking
could have both a high cost and be a threat to personnel safety.

There are also other stakeholders who may interact with the industry,
such as environmental authorities and health and safety authorities.
These did traditionally take manual samples, but are now starting
to use sensor network for real-time sampling either within or outside
the industry premises in order to perform continuous monitoring of
emissions or work environment. This is useful, and such continuous
monitoring can even be used by the company to optimise industrial
processes. There will however be privacy and confidentiality issues
with too fine-grained monitoring of such data. Information about problems
in production processes could for example affect the market value
of the company. This means that data access also for public authorities
will also require fine-grained access control as well as pre-processing
(for example averaging data) of the sensor data to avoid leaking detailed
sensitive information that can be used for example to infer how production
processes work.

Our survey for the oil and gas industry focused on equipment manufacturers
which wanted to monitor their equipment when used by oil rigs, mostly
for the drilling operation. This monitoring is part of a condition
based maintenance service. In this case, secrecy of data was a huge
problem for information sharing. The rig operator did not want to
share operations data with the equipment manufacturer, but the equipment
monitoring would reveal many parameters relevant to the operations
as the equipment manufacturer often delivered a complete drilling
package and monitored most equipment usage. The equipment manufacturer
on their side, would not let outsiders access to the monitoring data
as this revealed know-how about the equipment. These challenges was
not related only to IoT, but IoT can be said to be part of the scenario
as monitoring sensors get more advanced.

Availability was another challenge. Stable Internet connections with
good bandwidth could not be expected as drilling operations take place
all over the world, e.g. on ships where satellite is the only means
for communication.

Data quality was mostly a concern with regards to tampering, where
tampering could lead to false information. False information could
lead to wrong decisions, and as the cost rates for drilling are very
high, wrong decisions could have a high cost. Tampering was also a
concern for rig operators, as using IoT devices as a backdoor into
the control system could have fatal consequences for personnel and
environment safety and cost.

Despite all the challenges, the oil and gas industries are rapidly
moving towards the concept of Integrated Operations, where information
sharing and making decisions based on sensor data will have a big
role.

\emph{}

\section{Roadmap}

This section describe a roadmap for how secure information sharing
can be achieved in the industrial internet of things. A problem when
performing large-scale deployment of IIoT devices is that there are
standardisation efforts going on, however there is still a lack of
mature standard that have significant industrial adoption. There are
several reasons for this, for example that what consitutes a thing
varies widely from very simple purposed-build networked devices to
embedded devices running embedded or standard open source or commercial
operating systems. These devices have widely different capabilities,
which also affects what kind of services they can run to protect their
network environment and communication. Another challenge is that regulators
need to start focusing on the issue of insecure IIoT, and require
regulations and contracts for a certain minimum security standard
for IIoT devices. In parallel with this, there are big players such
as ARM, Intel and Google who have their own IoT device platforms as
well as cloud providers having their proprietary cloud interfaces
for these devices.

The next subsection describes one candidate IETF standard for securing
Internet-based IoT devices. Research, standardisation and industry
adoption by IIoT device vendors is probably the first step towards
increased security in IIoT. In parallel with this can existing organisation
already now use existing good practices for securing IIoT networks
in the short run. There are also some research initiatives that aim
at industrialising security solutions built around existing vulnerable
IIoT devices and SCADA systems using techniques such as software-defined
security. In the long run, standardised security solutions based on
existing industry standards for process control such as OPC-UA that
could consolidate the IIoT devices within the manufacturing plant
in a secure way with fine-grained access control.

\subsection{Security Considerations in the IP-based Internet of Things}

The Internet Engineering Task Force (IETF) has a work in progress
draft covering security considerations for IP-based IoT~\cite{garcia-morchon_security_2014}.
The draft examines the current state of the art, further possibilities,
and challenges in the security realm of IoT. 

An IoT device is referred to as a thing whose life cycle starts during
its manufacturing, and ends when it has been decommissioned by its
user. During the end of the manufacturing cycle, the thing has an
initial bootstrapping where it securely joins the IoT network at its
location. This also covers the initial authentication, authorisation
and configuration of necessary parameters for trusted operations in
the network. When the device is connected to the IoT network it is
considered operational until it needs maintenance, for example installing
a software update which is followed by a re-bootstrapping of said
device. This continues until the device is no longer in use and has
been decommissioned.

The life cycle presented is used as a base for identifying where possible
threats could happen. The threat analysis covers the following protocols:
HTTPS, 6LoWPAN, ANCP, DNS, SIP, IPv6, ND, and PANA. There are several
groups of threats considered which either compromises the thing itself
or the network as a whole: cloning, malicious substitution, eavesdropping,
man-in-the-middle, firmware replacement, extraction of security parameters,
routing attacks including sinkhole, black hole, privacy threats, and
Denial of Service. There is also a risk that things can be cloned
and sold for a cheaper price in the market by competitors. Untrusted
manufactures could also change the functionality of cloned devices
for example by adding a backdoor. Related to the cloning is malicious
substitution where one thing can be swapped with another ``copy''
of lower quality, which could lead to degraded functionality. Eavesdropping
attacks could happen during bootstrapping events before any secure
communication has been established, which can compromise the authenticity
and confidentiality of the communication channel. This phase may also
be vulnerable to man in the middle attacks. Firmware replacement attacks
can happen during a maintenance phase, where an attacker can exploit
the fact that the device is under update and install malicious firmware.

This draft standard presents the current state of art (2013), where
protocols such as ZigBee, BACNet, and DALI play the key roles, but
the trend is moving towards all-IP solutions. One of these solutions
is the 6LoWPAN working groups which focuses on transportation of IPv6
packets over IEEE 802.15.4 networks. For IP-based solutions there
is a plurality of security solutions to consider, and the draft identifies
and examines the following: IKEv2/IPsec, TLS/SSL, DTLS, HIP, PANA,
and EAP. One of the problems identified when using an IP based security
solution for IoT is that there are minor differences between IoT protocols
and regular Internet protocols. This could hamper end-to-end security
if communication relies on protocol translators between sender and
receiver. 

Five security profiles are defined in the draft standard ranging from
IoT devices with no security needs, home usage, managed home usage,
industrial usage, and advanced industrial usage. The industrial security
profile is where operation on devices relies on a central devices
for security, while advanced IIoT can also enable ad-hoc operations
between themselves or they can have more then one central control
device. Both of these profiles can have a network manager located
in a 6LoWPAN/CoAP network, which also handles the key management.
Under industrial usage, devices are required to be associated with
the network in a secure way the first time they are introduced. Broadcast
messaging should be secured with entity authentication (ID-CoAPMulticast).
Remote management is done through a backend manager which is in charge
of managing the different software installed or information exchanged
within the network. 

The draft identifies that a basic building block when considering
the next step towards a flexible and secure IoT for networks would
be DTLS, One promising implementation towards embedded development
is TinyDTLS which offers an open source implementation of the protocol
usable for resource constrained devices. Good solutions for bootstrapping
is still lacking, since there is a real need for good protocols that
resolves the initial authentication, authorisation and configuration.
Secure resource discovery security issues is still unclear, for example
on how to handle secure DNS and time synchronisation. Some vendors
have proposed proprietary extensions to handle this, such as the SmartAMM
protocol developed by Develco systems\footnote{SmartAMM: https://stateofgreen.com/en/profiles/develco-products/solutions/smartamm-makes-it-easier-to-monitor-private-households-electrical-appliances}.
The way security is layered where each layer take care of its own
need, might not be so feasible for a small device where resources
are tight. The draft argues that there should be more inter-connectivity
across these layers to be efficient and manage the whole security
from link to application instead of having multiple managers.

\subsection{Short-term Solution}

There are some basic principles that should be kept during the protection
of IIoT. To prevent and detect any malicious activity in the short
term, the following steps are recommended to be followed. The objective
of the short-term solution is that a vulnerable infrastructure can
be protected using a surrounding set of security tools based on existing
good practices such as firewalls, intrusion detection system, vulnerability
scanners etc. The PRECYSE security methodology, tools and architecture
is an example of solution based on existing and some new security
components that supports adding protection to a vulnerable critical
infrastructure this way~\cite{kippe_cyber-security_2014}. The PRECYSE
project did for example demonstrate adding protection of SCADA telecontrol
systems in the energy sector~\cite{yang_intrusion_2013}, as well
as vulnerable city traffic controllers~{[}In press{]}. The PRECYSE
architecture uses the concept of configurable security Domains and
Enclaves~\cite{rome_enclaves_2012}, where each Domain enforces a
given security policy for a given Enclave.

Other good practices that can be applied in the short run are:

\subsubsection{Network Segregation}

One approach that has been proposed for enforcing network segregation,
is adding surrounding security tools which effectively are able to
segregate and monitor the networks in order to provide higher security
awareness with identification of policy violations~\cite{kippe_cyber-security_2014}.
The objective then adding software-defined security solutions for
segregating the network, as well as monitoring the resulting network
Domains and Enclaves.

\subsubsection{Continuous Monitoring and Analysis}

Computer systems have become more and more complex which makes the
protection much more difficult. Due to the continuous rapid development
of sophisticated attacks and the previously known and unknown threats
and attack vectors, the most secure solution is to continuously analyse
the system behaviors and data. All computer system can be analysed
in several different ways.

\subsubsection{Log Analysis}

Most of all computer device and software such as network devices,
operating systems, applications and all manner of intelligent or programmable
devices document their activity by producing logs. Logs can be used
for auditing, or checking the compliance according to regulations
or trouble shooting. Logs are also good for forensic activities and
detecting intrusion attempts. Several attack types can be easily recognized
by log analysis such as attacks producing large amount of log entries
(e.g. brute forcing). Other types that have a definite attack pattern
can be detected easily as well. Host-based intrusion detection systems
typically support such log analysis.

\subsubsection{File Integrity Monitoring}

File integrity analysis is mainly for operating systems and software
for validating its integrity with some verification method. The most
frequently used verification method is the calculation of some kind
of cryptographic checksum (hash) which can be compared to a base value
or a list. Checksum verification can be used for identifying harmful
files (black listing) or it can be used for identifying allowed files
(white listings). The latter is obviously stricter and more secure
however from the point of view of functionality black listing is easier
to implement. Host based intrusion detection systems typically also
support file integrity monitoring.

\subsubsection{Network Traffic Analysis}

Network monitoring or network traffic analysis is needed for detecting
malicious activity by analyzing the network packets. Intrusion or
malicious activity recognition can be based on patterns or behavior
analysis. However sophisticated malware can hide the information in
covert channels, which can be so subtle that only pixels are changed
in a legitimate picture \cite{crysys_lab_duqu:_2011}. In that case,
network traffic analysis can only detect the suspicious destination
of the packet or the amount of network packets that are sent to the
destination (e.g. command and control server).

\subsubsection{Memory Dump Analysis}

Memory dump analysis is one of the best way of detecting unknown and
well known malware and malicious activity in the operating systems
memory. Volatility framework\footnote{The Volatility framework: http://www.volatilityfoundation.org/}
is able to analyse several type of memory dumps using advanced techniques.
Hidden processes as well as libraries loaded for malicious activity
can be detected, which facilitates the detection of sophisticated
intrusions into the system.

\subsubsection{Regular Malicious Activity Detecting Tools}

In addition to specific memory, network traffic and file analysis,
the usage of regular Anti-Virus (AV) and security products with up-to-date
attack pattern database and heuristic search methods is a must.

\subsubsection{Continuous Updating and Patching}

Continuous updating of the system and software (especially the 3rd
party software) is crucial from the security point of view. Unknown
software errors can provide the possibility of arbitrary code execution
on the operating system for the attacker. In \textquotedbl{}lucky\textquotedbl{}
cases a software error only leads to denial of service, which in itself
can have drastic effects on a critical infrastructures, since availability
typically is of paramount importance. Malicious attacks may be even
worse, since they may compromise the device without being detected,
and can be used as a bridgehead for further attacks into the critical
infrastructure as well as for industry espionage. It is important
to monitor security news sources and react on knowledge about new
vulnerabilities as quickly as possible.

\subsubsection{Regular Vulnerability Testing }

The security of a system is to a large extent determined by the design
of the system. Continuous monitoring should be used to detect any
malicious attempt, and vulnerability testing can draw the attention
to unknown errors. The vulnerability test can be related to the whole
system, or a specific component (e.g. software vulnerability test,
penetration test of a specific computer through the network, etc.
)

Vulnerability testing should be done at regular intervals since a
new analysis can reveal new threats. 

Vulnerability test can be done in terms of:
\begin{itemize}
\item Black box (the attacker has no access to the system and no previous
knowledge) 
\item Grey box (the attacker is a user of the system with restrictions) 
\item White box (the attacker has a good overview of the system, e.g. administrative
rights)
\end{itemize}
\emph{}

\subsubsection{Proxy solutions}

Using proxy solutions to build protection around legacy or vulnerable
solutions is a well-known technique for increasing the security. This
type of solutions can implement access control functionality, limit
the commands send to the protected device or network, perform inspection
and filtering, etc. In cases where a sub-set of data should be made
available for e.g. a supplier, the relevant data can be exported to
a DMZ using a trusted process, and thereby effectively remove the
need for giving the supplier access to the sensitive network. If needed
this trusted process can also implement functionality to reduce the
detail level of the exported data.

\subsection{Long-term Solution}

This section outlines possible long-term solutions for improving the
security of IIoT devices and facilitating necessary data access. It
is assumed that the long-term solution will include developing a security
gateway based on existing industry standards such as for example OPC-UA.
This would allow for integrating variety of IIoT devices and expose
them to external services according to a strict definition of need.
Significan research and development as well as standardisation and
industry adoption of these standards is however required before such
a solution will be successful.

\subsubsection{OPC-UA Managed Gateway for Controlled Information Access}

The web service mapping of OPC-UA supports the WS-Security standard,
and the native mapping maps these to similar cryptographic primitives.
OPC-UA supports its own service discovery, as well as using standard
service repositories such as LDAP or UDDI~\cite{hannelius_roadmap_2008}.
OPC-UA defines objects in terms of variables, methods and events.
This object model is mapped to the address space as nodes which are
interconnected by references. OPC-UA allows for interconnecting existing
OPC solutions using OPC UA wrappers and proxies~\cite{hannelius_roadmap_2008}.
Another method is utilising OPC UA gateways and adapters. 

An OPC UA wrapper is able to seamlessly integrate an OPC COM server~\cite{hannelius_roadmap_2008}.
The wrapper is responsible for handling endpoints and managing UA
encoding/decoding, security, transport and maping the COM server's
address space to UA~\cite{hannelius_roadmap_2008}. Data change call-backs
initiated by the COM server are returned as OPC UA Publish requests. 

An OPC UA proxy allows for conversion in the other direction, so that
OPC COM clients can communicate with an OPC UA server~\cite{hannelius_roadmap_2008}.
A problem with mapping using OPC UA proxies and wrappers is that it
is not able to map new concepts and technologies to old COM implementations~\cite{hannelius_roadmap_2008}.
Specifically, different profiles will be needed for mapping the OPC
Data Access (DA), Alarms and Events (AE) and Historical Data Access
(HDA) specifications to OPC UA, since these standards have different
semantics. Also, previous OPC specifications did not address security,
which means that functionality for managing confidentiality, integrity
and application authentication must be added on the OPC-UA side. Also,
performance, difference in transmission rate and latencies can be
an issue with such protocol conversion, depending on the real-time
requirements of the use case.

An OPC UA gateway is one possible solution for solving these issues
by integrating the different wrapping components, as well as adding
the necessary security functionality. The strong inner security model
of OPC UA facilitates hiding security sensitive processes from malicious
attacks, whilst still providing the necessary functionality for accessing
the underlying vulnerable COM-based infrastructure. Figure \ref{fig:OPC-UA-gateway}
illustrates at a high level how an UPC OA gateway can be extended
to support a service authorisation layer providing fine-grained access
to underlying OPC UA services based on an authorisation and privacy
policy. The gateway supports handling and converting messages between
OPC UA as well as the traditional COM-based infrastructure supporting
both OPC DA, A\&E and the HDA interfaces via the UA services API.
The gateway concept allows for supporting adapter plug-in modules
for adding new functionality that by default is not supported by the
standard conversion profiles.

\emph{}

\emph{}

\section{Related Work}

As the number of IoT devices proliferate, several research initiatives
focus on finding a general solution for the security of the IoT. Ukil
at al. proposed a solution for embedded security where the hardware
and its data aims to be secured~\cite{ukil_embedded_2011}. Also
a general solution is proposed by Cisco Security~\cite{frahim_securing_2015}
This is a framework that may be used in protocol and product development
as well as policy enforcement in operational environments.

In case of Industrial IoT, previous research mainly addresses threats
of IIoT. Sadeghi at al. gives an introduction to Industrial IoT systems,
the related security and privacy challenges, and an outlook on possible
solutions towards a holistic security framework for Industrial IoT
systems~\cite{sadeghi_security_2015}. Xu at al. summarises the current
state-of-the-art IoT in industries systematically~\cite{xu_internet_2014}.
Meltzer discusses security aspects of the Industrial Internet of Things
due to the explosion of IP-connected devices used in such areas as
control systems, manufacturing, utilities, and transportation~\cite{meltzer_securing_2015}.
Other studies focus on specific problems of IIoT such as the vulnerabilities
and risks in the industrial usage of wireless communication~\cite{plosz_security_2014}.
NSA provided a framework description for Assessing and Improving the
Security Posture of Industrial Control Systems~\cite{systems_and_network_analysis_center_framework_????}.

\section{Summary}

This article has proposed a roadmap for handling the problem of secure
information sharing with external vendors in an IIoT. It proposes
how IIoT should be secured both in the short term by applying existing
good practices in a structured manner, as well as utilising and extending
security toolsuites such as the PRECYSE architecture for protecting
vulnerable IIoT devices. In the long term we envisage that better
solutions will be needed, for example an OPC-UA gateway with support
for very fine-grained access control to data in IIoT devices. This
should be integrated with the organisation's own single-sign-on authentication
infrastructure, essentially providing the possibility for assigning
or revoking access to individual IIoT devices as well as providing
or denying access to certain data (individual XML elements or attributes)
within messages from these devices.

\section{Future Work}

Future work involves research on integrated solutions for protecting
vulnerable IIoT devices, for example by building software-defined
security solutions on top of existing frameworks such as the PRECYSE
architecture~\cite{kippe_cyber-security_2014}. Long-term research
could involve implementing an OPC-UA gateway with support for firewall
functionality as well as very finegrained access control, for example
based on the Reversible Anonymiser, which would allow for policy-controlled
access to individual data in the OPC-UA messages~\cite{ulltveit-moe_novel_2014}.

\section*{Acknowledgements}

This article was financed using research funding from VRI Agder in
Norway. Also thanks to the regional industry clusters Digin and Future
Robotics for initiating this research.

%\bibliographystyle{IEEEtran}
%\bibliography{bibliography,Mitt_bibliotek}

\begin{thebibliography}{10}
\providecommand{\url}[1]{#1}
\csname url@samestyle\endcsname
\providecommand{\newblock}{\relax}
\providecommand{\bibinfo}[2]{#2}
\providecommand{\BIBentrySTDinterwordspacing}{\spaceskip=0pt\relax}
\providecommand{\BIBentryALTinterwordstretchfactor}{4}
\providecommand{\BIBentryALTinterwordspacing}{\spaceskip=\fontdimen2\font plus
\BIBentryALTinterwordstretchfactor\fontdimen3\font minus
  \fontdimen4\font\relax}
\providecommand{\BIBforeignlanguage}[2]{{%
\expandafter\ifx\csname l@#1\endcsname\relax
\typeout{** WARNING: IEEEtran.bst: No hyphenation pattern has been}%
\typeout{** loaded for the language `#1'. Using the pattern for}%
\typeout{** the default language instead.}%
\else
\language=\csname l@#1\endcsname
\fi
#2}}
\providecommand{\BIBdecl}{\relax}
\BIBdecl

\bibitem{lee_recent_2013}
\BIBentryALTinterwordspacing
J.~Lee, E.~Lapira, B.~Bagheri, and H.-a. Kao, ``Recent advances and trends in
  predictive manufacturing systems in big data environment,''
  \emph{Manufacturing Letters}, vol.~1, no.~1, pp. 38--41, Oct. 2013. [Online].
  Available:
  \url{http://www.sciencedirect.com/science/article/pii/S2213846313000114}
\BIBentrySTDinterwordspacing

\bibitem{hannelius_roadmap_2008}
T.~Hannelius, M.~Salmenpera, and S.~Kuikka, ``Roadmap to adopting {OPC} {UA},''
  in \emph{6th {IEEE} {International} {Conference} on {Industrial}
  {Informatics}, 2008. {INDIN} 2008}, Jul. 2008, pp. 756--761.

\bibitem{vermesan_research_2014}
O.~Vermesan and P.~Friess, Eds., \emph{Internet of Things - From Research and
  Innovation to Market Deployment}.\hskip 1em plus 0.5em minus 0.4em\relax
  Niels Jernes Vej 10, 9220 Aalborg Ø: River Publishers, 2014.

\bibitem{kagermann_industrie_2011}
\BIBentryALTinterwordspacing
H.~Kagermann, L.~Wolf-Dieter, and W.~Wahlster, ``Industrie 4.0: {Mit} dem
  {Internet} der {Dinge} auf dem {Web} zur 4. industriellen {Revolution},''
  2011. [Online]. Available:
  \url{http://www.wolfgang-wahlster.de/wordpress/wp-content/uploads/Industrie_4_0_Mit_dem_Internet_der_Dinge_auf_dem_Weg_zur_vierten_industriellen_Revolution_2.pdf}
\BIBentrySTDinterwordspacing

\bibitem{hermann_design_2015}
\BIBentryALTinterwordspacing
M.~Hermann, T.~Pentek, and B.~Otto, ``Design {Principles} for {Industrie} 4.0
  {Scenarios}: {A} {Literature} {Review},'' 2015. [Online]. Available:
  \url{http://www.snom.mb.tu-dortmund.de/cms/de/forschung/Arbeitsberichte/Design-Principles-for-Industrie-4_0-Scenarios.pdf}
\BIBentrySTDinterwordspacing

\bibitem{crysys_lab_duqu:_2011}
\BIBentryALTinterwordspacing
C.~lab, ``Duqu: {A} {Stuxnet}-like malware found in the wild, technical
  report,'' 2011. [Online]. Available:
  \url{http://www.crysys.hu/publications/files/bencsathPBF11duqu.pdf}
\BIBentrySTDinterwordspacing

\bibitem{lab_duqu:_2011}
\BIBentryALTinterwordspacing
------, \emph{Duqu: {A} {Stuxnet}-like malware found in the wild, technical
  report}.\hskip 1em plus 0.5em minus 0.4em\relax Budapest University of
  Technology and Economics, 2011. [Online]. Available:
  \url{http://www.crysys.hu/publications/files/bencsathPBF11duqu.pdf}
\BIBentrySTDinterwordspacing

\bibitem{dini_considerations_2010}
G.~Dini and M.~Tiloca, ``Considerations on {Security} in {ZigBee} {Networks},''
  in \emph{2010 {IEEE} {International} {Conference} on {Sensor} {Networks},
  {Ubiquitous}, and {Trustworthy} {Computing} ({SUTC})}, Jun. 2010, pp. 58--65.

\bibitem{fouladi_honey_2013}
\BIBentryALTinterwordspacing
B.~Fouladi and S.~Ghanoun, ``Honey, {I}’m home!! - {Hacking} {Z}-{Wave}
  {Home} {Automation} {Systems},'' 2013. [Online]. Available:
  \url{https://www.blackhat.com/us-13/briefings.html}
\BIBentrySTDinterwordspacing

\bibitem{bezzi_secure_2010}
S.~Köpsell and P.~Švenda, ``Secure {Logging} of {Retained} {Data} for an
  {Anonymity} {Service},'' in \emph{Privacy and {Identity} {Management} for
  {Life}}, M.~Bezzi, P.~Duquenoy, S.~Fischer-Hübner, M.~Hansen, and G.~Zhang,
  Eds.\hskip 1em plus 0.5em minus 0.4em\relax Berlin, Heidelberg: Springer
  Berlin Heidelberg, 2010, vol. 320, pp. 284--298.

\bibitem{ulltveit-moe_novel_2014}
\BIBentryALTinterwordspacing
N.~Ulltveit-Moe and V.~Oleshchuk, ``A novel policy-driven reversible
  anonymisation scheme for {XML}-based services,'' \emph{Information Systems},
  2014. [Online]. Available:
  \url{http://www.sciencedirect.com/science/article/pii/S030643791400091X}
\BIBentrySTDinterwordspacing

\bibitem{papadimitriou_data_2011}
P.~Papadimitriou and H.~Garcia-Molina, ``Data {Leakage} {Detection},''
  \emph{IEEE Transactions on Knowledge and Data Engineering}, vol.~23, no.~1,
  pp. 51--63, Jan. 2011.

\bibitem{ulltveit-moe_measuring_2013}
\BIBentryALTinterwordspacing
N.~Ulltveit-Moe and V.~Oleshchuk, ``Measuring {Privacy} {Leakage} for {IDS}
  {Rules},'' \emph{arXiv:1308.5421 [cs, math]}, Aug. 2013, arXiv: 1308.5421.
  [Online]. Available: \url{http://arxiv.org/abs/1308.5421}
\BIBentrySTDinterwordspacing

\bibitem{thomalla_ontologie-basierte_2014}
\BIBentryALTinterwordspacing
C.~Thomalla, ``Ontologie-basierte erkennung,'' in \emph{{visIT}
  {IT}-{Sicherheit} für die {Produktion}}.\hskip 1em plus 0.5em minus
  0.4em\relax Fraunhofer IOSB, 2014, vol.~15, no. ISSN 1616-8240. [Online].
  Available:
  \url{https://www.iosb.fraunhofer.de/servlet/is/49961/visIT%20%5BIT-Sicherheit%20in%20der%20Produktion%5D.pdf?command=downloadContent&filename=visIT%20%5BIT-Sicherheit%20in%20der%20Produktion%5D.pdf}
\BIBentrySTDinterwordspacing

\bibitem{moses_oasis_2005}
\BIBentryALTinterwordspacing
T.~e. Moses, \emph{{OASIS} {eXtensible} {Access} {Control} {Markup} {Language}
  ({XACML}) {Version} 2.0}, 2005. [Online]. Available:
  \url{http://docs.oasis-open.org/xacml/2.0/access_control-xacml-2.0-core-spec-os.pdf}
\BIBentrySTDinterwordspacing

\bibitem{uslander_designing_2010}
\BIBentryALTinterwordspacing
T.~Usländer, P.~Jacques, I.~Simonis, and K.~Watson, ``Designing
  {Environmental} {Software} {Applications} {Based} {Upon} an {Open} {Sensor}
  {Service} {Architecture},'' \emph{Environ. Model. Softw.}, vol.~25, no.~9,
  pp. 977--987, Sep. 2010. [Online]. Available:
  \url{http://dx.doi.org/10.1016/j.envsoft.2010.03.013}
\BIBentrySTDinterwordspacing

\bibitem{open_geospatial_consortium_rm_oa_2007}
\BIBentryALTinterwordspacing
{Open Geospatial Consortium}, ``Reference model for the orchestra architecture
  (rm-oa) v2,'' 2007. [Online]. Available:
  \url{http://portal.opengeospatial.org/files/?artifact_id=20300&passcode=5492ay9tzwprgwbytk9a}
\BIBentrySTDinterwordspacing

\bibitem{open_geospatial_consortium_geodrm_2006}
\BIBentryALTinterwordspacing
------, ``Geospatial digital rights management reference model {(GeoDRM RM)},''
  2006. [Online]. Available:
  \url{http://portal.opengeospatial.org/files/?artifact_id=14085&passcode=5492ay9tzwprgwbytk9a}
\BIBentrySTDinterwordspacing

\bibitem{ed_ogc_2007}
\BIBentryALTinterwordspacing
A.~M. (ed), \emph{{OGC} 07-026r2 {Geospatial} {eXtensible} {Access} {Control}
  {Markup} {Language} ({GeoXACML}) version 1.0}.\hskip 1em plus 0.5em minus
  0.4em\relax Open Geospatial Consortium, Inc., 2007. [Online]. Available:
  \url{http://portal.opengeospatial.org/files/?artifact_id=25218}
\BIBentrySTDinterwordspacing

\bibitem{matheus_geospatial_2008}
A.~Matheus and J.~Herrmann, ``Geospatial extensible access control markup
  language (geoxacml),'' \emph{Open Geospatial Consortium Inc}, 2008.

\bibitem{ed_ogc_2002}
\BIBentryALTinterwordspacing
S.~C. e.~a. (ed), \emph{{OGC} 02-023r4 {OpenGIS} {Geography} {Markup}
  {Language} ({GML}) {Encoding} {Specification} {Version} 3.00}.\hskip 1em plus
  0.5em minus 0.4em\relax Open Geospatial Consortium, Inc., 2002. [Online].
  Available: \url{https://portal.opengeospatial.org/files/?artifact_id=7174}
\BIBentrySTDinterwordspacing

\bibitem{nergaard_scratch-based_2015}
H.~Nergaard, N.~Ulltveit-Moe, and T.~Gjøsæter, ``A {Scratch}-based
  {Graphical} {Policy} {Editor} for {XACML},'' in \emph{{ICISSP} 2015
  {Proceedings} of the 1st {International} {Conference} on {Information}
  {Systems} {Security} and {Privacy} {ESEO}, {Angers}, {Loire} {Valley},
  {France}}.\hskip 1em plus 0.5em minus 0.4em\relax Scitepress, 2015, pp.
  182--191.

\bibitem{gjosaeter_security_2014}
\BIBentryALTinterwordspacing
T.~Gjøsæter, N.~Ulltveit-Moe, M.~L. Kolhe, R.~H. Jacobsen, and E.~S.~M.
  Ebeid, ``Security and {Privacy} in the {SEMIAH} {Home} {Energy} {Management}
  {System},'' 2014. [Online]. Available:
  \url{https://www.google.no/url?sa=t&rct=j&q=&esrc=s&source=web&cd=1&cad=rja&uact=8&ved=0ahUKEwjvuvqZ6NnJAhXIjSwKHTpzCmUQFggfMAA&url=http%3A%2F%2Fsemiah.eu%2Fwp-content%2Fuploads%2F2014%2F10%2F2014-Security-and-Privacy-in-the-SEMIAH.pdf&usg=AFQjCNFGGtwXP69Dgq08_2TiNsHWZlTrEQ&sig2=euoqjVZieoTj1CsWSAm2Jw}
\BIBentrySTDinterwordspacing

\bibitem{goodin_malicious_2015}
\BIBentryALTinterwordspacing
D.~Goodin, ``Malicious {Cisco} router backdoor found on 79 more devices, 25 in
  the {US},'' 2015. [Online]. Available:
  \url{http://arstechnica.com/security/2015/09/malicious-cisco-router-backdoor-found-on-79-more-devices-25-in-the-us/}
\BIBentrySTDinterwordspacing

\bibitem{shamir_how_1979}
A.~Shamir, ``How to share a secret,'' \emph{Commun. ACM}, vol.~22, no.~11, pp.
  612--613, 1979.

\bibitem{Jupiterresearch}
\BIBentryALTinterwordspacing
{Jupiter Research}, ``{ The Internet of Things: Consumer, Industrial \& Public
  Services 2015-2020},'' 2012. [Online]. Available:
  \url{http://www.juniperresearch.com/researchstore/key-vertical-markets/internet-of-things/consumer-industrial-public-services?utm_source=juniperpr&utm_medium=email&utm_campaign=iot15pr1%20[2]%20http://www.go-gulf.com/blog/cyber-crime/}
\BIBentrySTDinterwordspacing

\bibitem{Cybercrimestatistics}
\BIBentryALTinterwordspacing
{Go-Gulf}, ``{CYBER CRIME STATISTICS AND TRENDS [INFOGRAPHIC]},'' 2015.
  [Online]. Available: \url{http://www.go-gulf.com/blog/cyber-crime/}
\BIBentrySTDinterwordspacing

\bibitem{McAffeeThreatReport}
\BIBentryALTinterwordspacing
{Mcaffee}, ``{Mcaffee - Threat Report 2015 August},'' 2015. [Online].
  Available:
  \url{http://www.mcafee.com/us/resources/reports/rp-quarterly-threats-aug-2015.pdf}
\BIBentrySTDinterwordspacing

\bibitem{AnatomyofAttack}
\BIBentryALTinterwordspacing
{TrapX Security}, ``{Anatomy of an attack - The Internet of Things (IoT) },''
  2015. [Online]. Available:
  \url{http://www.mcafee.com/us/resources/reports/rp-quarterly-threats-aug-2015.pdf}
\BIBentrySTDinterwordspacing

\bibitem{HeartBleed}
\BIBentryALTinterwordspacing
{US-CERT}, ``{OpenSSL 'Heartbleed' vulnerability (CVE-2014-0160) },'' 2014.
  [Online]. Available: \url{https://www.us-cert.gov/ncas/alerts/TA14-098A}
\BIBentrySTDinterwordspacing

\bibitem{LZOVulnerability}
\BIBentryALTinterwordspacing
{Lab Mouse Security}, ``{Raising Lazarus - The 20 Year Old Bug that Went to
  Mars },'' 2014. [Online]. Available:
  \url{http://blog.securitymouse.com/2014/06/raising-lazarus-20-year-old-bug-that.html}
\BIBentrySTDinterwordspacing

\bibitem{CVEDetails}
\BIBentryALTinterwordspacing
{CVE Details webpage}, ``{CVE Details - The ultimate security vulnerability
  datasource },'' 2015. [Online]. Available: \url{http://www.cvedetails.com}
\BIBentrySTDinterwordspacing

\bibitem{KasperskySecBullet2013}
\BIBentryALTinterwordspacing
{Kasprsky Lab}, ``{Kaspersky Security Bulletin 2013. Corporate threats },''
  2013. [Online]. Available:
  \url{https://securelist.com/analysis/kaspersky-security-bulletin/58262/kaspersky-security-bulletin-2013-corporate-threats}
\BIBentrySTDinterwordspacing

\bibitem{RifleHack}
\BIBentryALTinterwordspacing
{Sandvik and Auger}, ``{When IoT Attacks: Hacking A Linux-Powered Rifle },''
  2015. [Online]. Available:
  \url{https://www.blackhat.com/docs/us-15/materials/us-15-Sandvik-When-IoT-Attacks-Hacking-A-Linux-Powered-Rifle.pdf}
\BIBentrySTDinterwordspacing

\bibitem{VectraThreat}
\BIBentryALTinterwordspacing
{Vectra Threat Labs}, ``{Belkin F9K1111 V1.04.10 Firmware Analysis },'' 2015.
  [Online]. Available:
  \url{http://blog.vectranetworks.com/blog/belkin-analysis}
\BIBentrySTDinterwordspacing

\bibitem{BabyMonitor}
\BIBentryALTinterwordspacing
{Rapid7}, ``{HACKING IoT: A Case Study on Baby Monitor Exposures and
  Vulnerabilities },'' 2015. [Online]. Available:
  \url{https://www.rapid7.com/docs/Hacking-IoT-A-Case-Study-on-Baby-Monitor-Exposures-and-Vulnerabilities.pdf}
\BIBentrySTDinterwordspacing

\bibitem{Slammer}
\BIBentryALTinterwordspacing
{David Moore}, ``{Inside the Slammer Worm },'' 2003. [Online]. Available:
  \url{http://www.icsi.berkeley.edu/pubs/networking/insidetheslammerworm03.pdf}
\BIBentrySTDinterwordspacing

\bibitem{Confiker}
\BIBentryALTinterwordspacing
{Dave Piscitello}, ``{Conficker Summary and Review},'' 2010. [Online].
  Available:
  \url{https://www.icann.org/en/system/files/files/conficker-summary-review-07may10-en.pdf}
\BIBentrySTDinterwordspacing

\bibitem{Stuxnet}
\BIBentryALTinterwordspacing
{Ralph Langner}, ``{To Kill a Centrifuge - A Technical Analysis of What
  Stuxnet’s Creators Tried to Achieve },'' 2013. [Online]. Available:
  \url{http://www.langner.com/en/wp-content/uploads/2013/11/To-kill-a-centrifuge.pdf}
\BIBentrySTDinterwordspacing

\bibitem{StuxnetTwin}
\BIBentryALTinterwordspacing
------, ``{Stuxnet’s Secret Twin },'' 2013. [Online]. Available:
  \url{http://foreignpolicy.com/2013/11/19/stuxnets-secret-twin/}
\BIBentrySTDinterwordspacing

\bibitem{Flame}
\BIBentryALTinterwordspacing
{Crysis Lab}, ``{sKyWIper (a.k.a. Flame a.k.a. Flamer): A complex malware for
  targeted attacks },'' 2012. [Online]. Available:
  \url{http://www.crysys.hu/skywiper/skywiper.pdf}
\BIBentrySTDinterwordspacing

\bibitem{Havex}
\BIBentryALTinterwordspacing
{F-Secure Labs}, ``{Havex Hunts For ICS/SCADA Systems},'' 2014. [Online].
  Available: \url{https://www.f-secure.com/weblog/archives/00002718.html}
\BIBentrySTDinterwordspacing

\bibitem{garcia-morchon_security_2014}
\BIBentryALTinterwordspacing
O.~Garcia-Morchon, S.~Kumar, {Philips Research}, S.~Keoh, {University of
  Glasgow}, R.~Hummen, {RWTH Aachen}, R.~Struik, and {Struik Consultancy},
  ``Security {Considerations} in the {IP}-based {Internet} of {Things},'' 2014.
  [Online]. Available:
  \url{https://tools.ietf.org/html/draft-garcia-core-security-06}
\BIBentrySTDinterwordspacing

\bibitem{kippe_cyber-security_2014}
\BIBentryALTinterwordspacing
J.~Kippe, ``Cyber-security in kritischen infrastrukturen,'' in \emph{{visIT}
  {IT}-{Sicherheit} für die {Produktion}}.\hskip 1em plus 0.5em minus
  0.4em\relax Fraunhofer IOSB, 2014, vol.~15, no. ISSN 1616-8240. [Online].
  Available:
  \url{https://www.iosb.fraunhofer.de/servlet/is/49961/visIT%20%5BIT-Sicherheit%20in%20der%20Produktion%5D.pdf?command=downloadContent&filename=visIT%20%5BIT-Sicherheit%20in%20der%20Produktion%5D.pdf}
\BIBentrySTDinterwordspacing

\bibitem{yang_intrusion_2013}
Y.~Yang, K.~McLaughlin, T.~Littler, S.~Sezer, B.~Pranggono, and H.~Wang,
  ``Intrusion {Detection} {System} for {IEC} 60870-5-104 based {SCADA}
  networks,'' in \emph{2013 {IEEE} {Power} and {Energy} {Society} {General}
  {Meeting} ({PES})}, Jul. 2013, pp. 1--5.

\bibitem{rome_enclaves_2012}
\BIBentryALTinterwordspacing
J.~A. Rome, ``Enclaves and {Collaborative} {Domains},'' 2012. [Online].
  Available:
  \url{http://jamesrome.home.comcast.net/~jamesrome/security/EnclavesAndCollaborativeDomains.pdf}
\BIBentrySTDinterwordspacing

\bibitem{ukil_embedded_2011}
A.~Ukil, J.~Sen, and S.~Koilakonda, ``Embedded security for {Internet} of
  {Things},'' in \emph{2011 2nd {National} {Conference} on {Emerging} {Trends}
  and {Applications} in {Computer} {Science} ({NCETACS})}, Mar. 2011, pp. 1--6.

\bibitem{frahim_securing_2015}
\BIBentryALTinterwordspacing
J.~Frahim, C.~Pignataro, J.~Apcar, and M.~Morrow, ``Securing the {Internet} of
  {Things}: {A} {Proposed} {Framework},'' 2015. [Online]. Available:
  \url{http://www.cisco.com/web/about/security/intelligence/iot_framework.html}
\BIBentrySTDinterwordspacing

\bibitem{sadeghi_security_2015}
\BIBentryALTinterwordspacing
A.-R. Sadeghi, C.~Wachsmann, and M.~Waidner, ``Security and {Privacy}
  {Challenges} in {Industrial} {Internet} of {Things},'' in \emph{Proceedings
  of the 52Nd {Annual} {Design} {Automation} {Conference}}, ser. {DAC}
  '15.\hskip 1em plus 0.5em minus 0.4em\relax New York, NY, USA: ACM, 2015, pp.
  54:1--54:6. [Online]. Available:
  \url{http://doi.acm.org/10.1145/2744769.2747942}
\BIBentrySTDinterwordspacing

\bibitem{xu_internet_2014}
L.~D. Xu, W.~He, and S.~Li, ``Internet of {Things} in {Industries}: {A}
  {Survey},'' \emph{IEEE Transactions on Industrial Informatics}, vol.~10,
  no.~4, pp. 2233--2243, Nov. 2014.

\bibitem{meltzer_securing_2015}
\BIBentryALTinterwordspacing
D.~Meltzer, ``Securing the {Industrial} {Internet} of {Things},'' 2015.
  [Online]. Available:
  \url{https://www.issa.org/resource/resmgr/journalpdfs/feature0615.pdf}
\BIBentrySTDinterwordspacing

\bibitem{plosz_security_2014}
S.~Plosz, A.~Farshad, M.~Tauber, C.~Lesjak, T.~Ruprechter, and N.~Pereira,
  ``Security vulnerabilities and risks in industrial usage of wireless
  communication,'' in \emph{2014 {IEEE} {Emerging} {Technology} and {Factory}
  {Automation} ({ETFA})}, Sep. 2014, pp. 1--8.

\bibitem{systems_and_network_analysis_center_framework_????}
\BIBentryALTinterwordspacing
{Systems and Network Analysis Center}, ``A {Framework} for {Assessing} and
  {Improving} the {Security} {Posture} of {Industrial} {Control} {Systems}
  ({ICS}).'' [Online]. Available:
  \url{https://www.nsa.gov/ia/_files/ics/ics_fact_sheet.pdf}
\BIBentrySTDinterwordspacing

\end{thebibliography}


\end{document}